\documentclass[iop,apj,appendixfloats]{emulateapj}
\usepackage{apjfonts,amsmath}

\gdef\msun{$M_{\odot}$}
\lefthead{van Dokkum et al.}
\righthead{}
\slugcomment{Accepted for publication in ApJ Letters}
\begin{document}

\title{First Results from the Dragonfly Telephoto Array:
the Apparent Lack of a Stellar Halo in
the Massive Spiral Galaxy M101}

\author{Pieter G.\ van Dokkum\altaffilmark{1},
Roberto Abraham\altaffilmark{2},
Allison Merritt\altaffilmark{1}
}

\altaffiltext{1}
{Department of Astronomy, Yale University, New Haven, CT 06511, USA}
\altaffiltext{2}
{Department of Astronomy \& Astrophysics, University of Toronto,
50 St.\ George St.,  Toronto, ON M5S 3H8, Canada}

\begin{abstract}
We use a new telescope concept, the Dragonfly
Telephoto Array, to study the low surface brightness
outskirts of the spiral
galaxy M101. The radial surface brightness profile is measured down to
$\mu_g \sim 32$\,mag\,arcsec$^{-2}$, a depth that approaches the sensitivity of
star count studies in the Local Group. We convert surface brightness
to surface mass density using the radial $g-r$ color profile. The mass
density profile shows no significant upturn at large radius and is
well-approximated by a simple bulge + disk model out to
$R = 70$\,kpc, corresponding to 18 disk scale lengths.
Fitting  a bulge + disk + halo model
we find that the best-fitting halo mass
$M_{\rm halo} = 1.7_{-1.7}^{+3.4} \times 10^8$\,\msun.
The total stellar mass of M101 is $M_{\rm tot,*}=
5.3_{-1.3}^{+1.7} \times 10^{10}$\,\msun, and
we infer that the halo mass fraction $f_{\rm halo}=M_{\rm halo}/M_{\rm tot,*}=
0.003^{+0.006}_{-0.003}$. This mass fraction is lower than
that of the Milky Way ($f_{\rm halo}\sim 0.02$) and M31 ($f_{\rm halo}
\sim 0.04$).
All three galaxies fall below the $f_{\rm halo}$ -- $M_{\rm tot,*}$
relation predicted by recent cosmological
simulations that trace the light of disrupted satellites, with M101's halo
mass a factor of $\sim 10$ below the median expectation.
However, the predicted  scatter in this relation is large,
and more galaxies
are needed to better quantify this possible
tension with galaxy formation models. Dragonfly is well suited
for this project: as integrated-light
surface brightness is
independent of distance, large numbers of galaxies can be studied
in a uniform way.

\end{abstract}


\section{Introduction}

Star counts in the direction of the Andromeda galaxy (M31)
have shown that it is embedded in a large, complex stellar halo
with significant substructure ({Ibata} {et~al.} 2001; {McConnachie} {et~al.} 2009).
Such  halos are thought to be
comprised of the debris of shredded satellite
galaxies ({Searle} \& {Zinn} 1978; {Newberg} {et~al.} 2002; {McConnachie} {et~al.} 2009),
and their existence around luminous spiral galaxies is a central
prediction of galaxy formation
models
({Bullock} \& {Johnston} 2005; {Abadi}, {Navarro}, \& {Steinmetz} 2006; {Purcell}, {Bullock}, \&  {Zentner} 2008; {Johnston} {et~al.} 2008; {Cooper} {et~al.} 2010; {Mart{\'{\i}}nez-Delgado} {et~al.} 2010; {Cooper} {et~al.} 2013).

Testing these predictions for large samples of galaxies is difficult,
as star count studies are limited to relatively small distances
({Barker} {et~al.} 2009; {Tanaka} {et~al.} 2011). A solution to this problem
is to study tidal features and stellar
halos in integrated light rather than star counts,
as the integrated surface brightness is independent
of distance\footnote{This is only true for
distances where the $(1+z)^4$ cosmological surface brightness
dimming is unimportant.}
(see, e.g., {Mihos} {et~al.} 2005; {van Dokkum} 2005; {Tal} {et~al.} 2009; {Mart{\'{\i}}nez-Delgado} {et~al.} 2010; {Atkinson}, {Abraham}, \&  {Ferguson} 2013).
However, conventional reflecting
telescopes cannot reliably observe low surface brightness emission
below $\mu_B \sim 29$\,mag\,arcsec$^{-2}$ due to 
systematic errors in flat fielding and
the complex point spread functions of stars (see {Slater}, {Harding}, \& {Mihos} 2009, for an in-depth
discussion of these issues).

The Dragonfly Telephoto Array ({Abraham} \& {van Dokkum} 2013) is a new class of
telescope that is optimized for detecting spatially-extended low
surface brightness emission.
The prototype Dragonfly telescope consists of
eight Canon EF 400\,mm f/2.8L IS II USM telephoto lenses on a common mount.
Telephoto lenses have no central obstruction and are optimally baffled.
Furthermore, the specific lenses used in the array are of superb optical
quality, partially owing to nano-fabricated sub-wavelength corrugations
on their anti-reflecting coatings.
As we show in {Abraham} \& {van Dokkum} (2013) the Dragonfly point spread function
has an order of magnitude less scattered light
than the best reflecting telescopes.
By combining eight lenses that image the same area of sky
we built a ``compound eye'' that acts as a
40\,cm f/1.0 refractor.
In its default configuration, four of the lenses are equipped with
SDSS $g$-band filters and four with SDSS $r$ filters. 
The pixel scale is $2.8''$ and the angular
field covered by each camera is $2.6^{\circ}\times 1.9^{\circ}$.

In this {\em Letter} we present the first results from Dragonfly: a study
of the stellar halo of the
well-known galaxy M101 (a.k.a.\ the Pinwheel
Galaxy). With a distance of
7\,Mpc ({Lee} \& {Jang} 2012), an absolute magnitude $r=-21.5$,
and a stellar mass of $\approx 5 \times 10^{10}$\,\msun\
(see \S\,4),
M101 is one of the nearest massive spiral
galaxies. 
M101 has been the subject of many detailed studies
(e.g., {van der Hulst} \& {Sancisi} 1988; {Kenney}, {Scoville}, \& {Wilson} 1991; {Kennicutt}, {Bresolin}, \&  {Garnett} 2003; {Mihos} {et~al.} 2012),
including one of the deepest photometric
investigations of the outskirts of galaxies beyond the Local Group done
so far ({Mihos} {et~al.} 2013).

\begin{figure*}[htbp]
\epsfxsize=18.0cm
\epsffile[0 0 975 707]{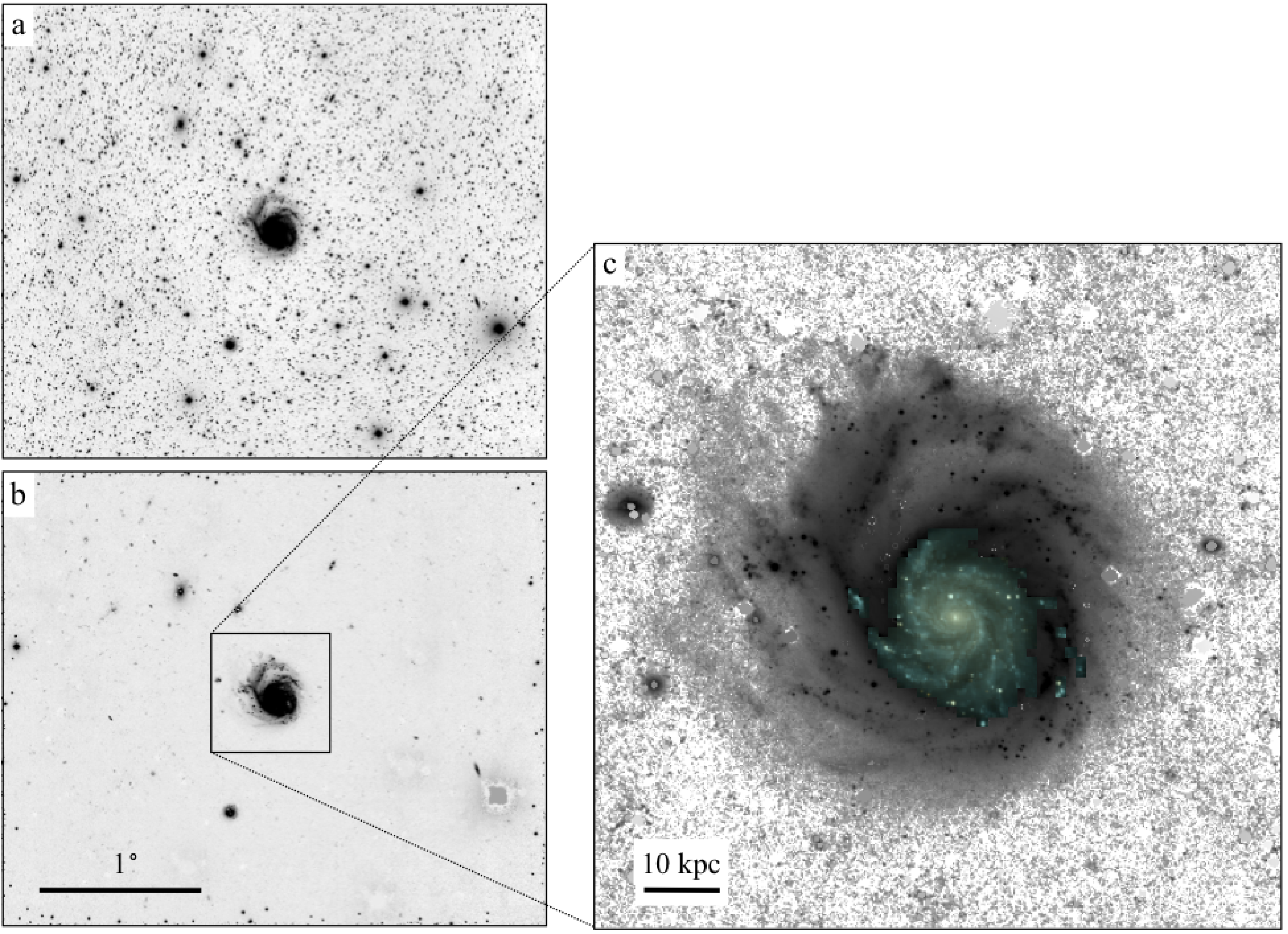}
\caption{\small
{\em a)} Dragonfly $g$-band image of the $3.33^{\circ} \times
2.78^{\circ}$ area centered on the galaxy M101. North is up
and East is to the left. {\em b)} The same image, after 
subtraction of stars and a model for large scale ($\gtrsim 1^{\circ}$)
background structure. Owing to the excellent PSF of Dragonfly, stars
as bright as $\sim 6^{\rm th}$ magnitude affect only a relatively
small number of pixels, and can be subtracted.
{\em c)} The $44' \times 44'$ area around M101 at high contrast. The
faint spiral arms on the East side of M101
(Mihos et al.\ 2013) have a surface brightness
of $\mu_g \sim 29$\,mag\,arcsec$^{-2}$.
The
color image in the center was created from the $g$ and $r$ exposures.
\label{full.fig}}
\end{figure*}

\section{Observations and Reduction}

Dragonfly is located at the New Mexico Skies observatory
in Mayhill, NM. It
is robotic and
operates in a semi-autonomous way.
A total of 35 hours of
observations were obtained in 13 nights in May and June 2013.
A typical M101
observing sequence consisted of nine dithered 600\,s exposures with
all eight cameras. The size of the dither box was typically
$50' \times 50'$. As we
have four slightly offset cameras per filter this provides
36 independent lines of sight
per filter for each 1.5 hr of observing time. Sky flats
were taken at the beginning of
each night and darks were taken throughout each night.

The data reduction followed standard procedures for imaging data, taking
care to preserve the large scale faint structure in the images. After
initial dark subtraction and flat fielding with night-specific
calibration frames a low-order illumination correction
was applied, created from a large number of
dithered observations over many nights.
Each frame was also
corrected for the 1\,\% -- 2\,\% gradient in the night
sky emission across the field-of-view
(see {Garstang} 1989), by subtracting a tilted plane.
Combined $g$ and $r$ images
were created for each night, using optimal weighting.
The 13 images of all nights were combined for each filter,
again using optimal weighting. 
The reduced, combined $g$-band image
is shown in Fig.\ \ref{full.fig}a.


The background in the images shows large scale variation at a level of
$\approx 0.2$\,\%
(peak-to-peak) over the $3.3^{\circ}\times 2.8^{\circ}$ field (see Fig.\
\ref{full.fig}a). As this background is independent of
camera orientation and variations in the dither pattern it is most likely
Galactic cirrus emission,\footnote{Some
independent support for this comes from the IRAS $100\,\mu$m image
of the M101 field, which shows a broadly similar morphology
(e.g., {Zagury}, {Boulanger}, \&  {Banchet} 1999).} at
levels of $\gtrsim 30$\,mag\,arcsec$^{-2}$.
This large scale background was removed by fitting a third-order polynomial
to a background image determined with SExtractor, aggressively masking
M101 and other objects in the field. In the analysis of the
surface brightness profile of M101 the average subtracted background
value in a particular radial bin was
added in quadrature to the uncertainty in the measured
surface brightness in that bin.

Stars were removed by modeling their spatially-varying point spread function
(PSF). First, large numbers of bright but unsaturated stars were used to
construct average PSFs in image sections. These PSFs were then
interpolated so that a PSF can be constructed for any location in
the image. Next, the wings of the PSF were modeled by
averaging saturated stars over the entire image. Care was
taken to mask neighboring stars in an iterative way when
doing the averaging, both when
determining the spatially-dependent inner parts of the PSF and when
constructing its wings. The background- and star-subtracted
$g$-band image is shown in Fig.\ \ref{full.fig}b.

\section{The Surface Brightness Profile of M101}

The central $44' \times 44'$ of
the star-subtracted $g$-band image, binned to a scale
of $6''$ per pixel, is shown in Fig.\ \ref{full.fig}c.
The image shows
the outer spiral arms of M101, including the faint extensions to the East
that were first identified by {Mihos} {et~al.} (2013).
This emission, which
corresponds to spiral structure seen in neutral Hydrogen
emission ({Walter} {et~al.} 2008; {Mihos} {et~al.} 2012, 2013), has a 
surface brightness of $\mu_g \sim 29$\,mag\,arcsec$^{-2}$.

\begin{figure}[htbp]
\epsfxsize=8.2cm
\epsffile[40 300 339 717]{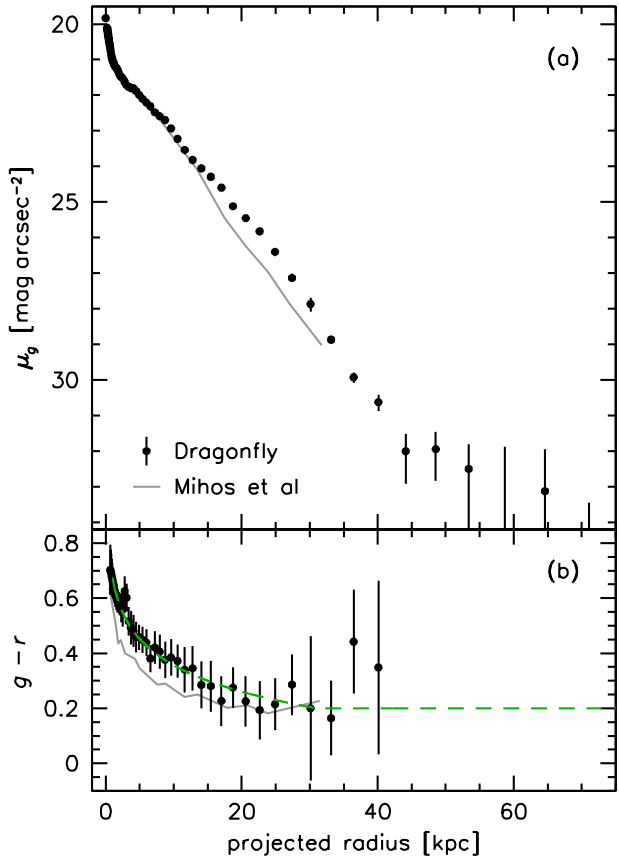}
\caption{\small
{\em a)} Radial $g$-band surface brightness profile of M101. The data reach
$\mu_g \approx 32$\,mag\,arcsec$^{-2}$. The grey line shows the
profile of Mihos et al.\ (2013), converted to the $g$ band.
{\em b)} Color profile derived from the $g$ and $r$ images.  The broken
green line is a constrained fit to the profile (Eq.\ 2).
\label{sbprof.fig}}
\end{figure}

We do not see coherent
faint emission at larger radii. The surface brightness of M101 falls
off rapidly outside of the area defined by the spiral arms, and we see
no evidence for an extended stellar halo or
features such as M31's ``giant stream'' ({Ibata} {et~al.} 2001).
We quantify
this visual impression with 
the projected surface brightness profile, shown in Fig.\ \ref{sbprof.fig}a.
The profile was determined by averaging the flux in circular
annuli at increasing distance from the center of the galaxy.
The profile reaches $\mu_g \sim 32$\,mag\,arcsec$^{-2}$ at
$R\approx 40$\,kpc, and there is no evidence for an upturn that might
have indicated a regime where light from
the stellar halo dominates over that of the
disk. The $g-r$ color profile is shown in Fig.\ \ref{sbprof.fig}b. The galaxy
becomes progressively bluer at larger radii. 

The grey line in Fig.\ \ref{sbprof.fig}a shows the surface brightness
profile measured by {Mihos} {et~al.} (2013), converted from $\mu_B$ to $\mu_g$ using
their $B-V$ profile and
Eq.\ 23 in {Fukugita} {et~al.} (1996). Within $R = 15$\,kpc the datasets agree
to $\lesssim 0.05$\,mag. At $R=20-30$\,kpc there is a discrepancy,
which is caused by a difference in methodology: {Mihos} {et~al.} (2013) determined
the median flux in each radial bin (C.\ Mihos, priv.\ comm.),
whereas we use the mean. The mean and median are different at
radial distances where the NE spiral arm is prominent. 
The grey line in
the bottom panel of Fig.\ \ref{sbprof.fig} shows
the $B-V$ profile of {Mihos} {et~al.} (2013), converted to $g-r$
({Fukugita} {et~al.} 1996). The profiles are offset by $\Delta(g-r) = 0.08\pm 0.05$,
where the error bar reflects the uncertainty in our zeropoint determinations
only. Assuming a similar uncertainty in the {Mihos} {et~al.} (2013) zeropoints
and/or their conversions to standard filters, the difference is not significant.

\begin{figure*}[htbp]
\epsfxsize=18cm
\epsffile[18 420 592 718]{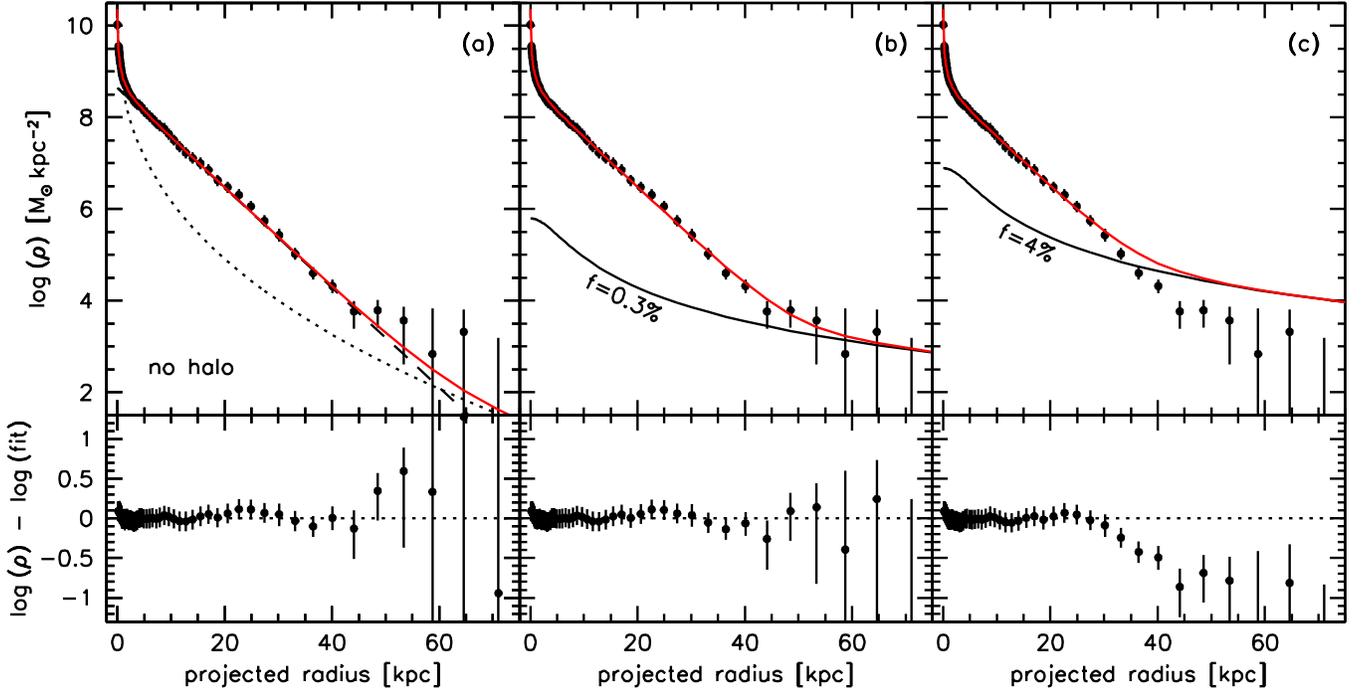}
\caption{\small
{\em a)} Mass density profile of M101. The red line is a bulge + disk
fit to the profile, with the individual components indicated by broken
lines. This fit provides an adequate description of the full profile,
as shown by the residuals in the bottom panel. {\em b)} Best-fitting
bulge + disk + halo fit. The best fit is obtained
for a stellar halo contributing $0.3^{+0.6}_{-0.3}$\,\% of the
total mass. {\em c)} Fit with a halo contributing 4\,\% of the mass,
the same as the M31 halo. This is a poor fit:
the halo of M101 is much less prominent than that of M31.
\label{massdens.fig}}
\end{figure*}

\section{Mass Density Profile and Constraints on the M101 Stellar
Halo}

\subsection{Construction of the Mass Density Profile}

We quantify  the contribution of the stellar halo to the total
mass of M101 by fitting the radial profile of M101.
We first convert the observed
surface brightness profile to a radial mass density
profile. This step is important as M101 has a strong color gradient
(Fig.\ 2b),
and the mass-to-light ($M/L$) ratio correlates
with color ({Bell} \& {de Jong} 2001). We use the following relation between
surface brightness and stellar surface density density:
\begin{equation}
\label{mlcol.eq}
\log(\rho)= -0.4(\mu_g -29.23) + 1.49(g-r) + 4.58,
\end{equation}
with $\mu_g$ in mag\,arcsec$^{-2}$ and $\rho$ in $M_{\odot}$\,kpc$^{-2}$.
Equation \ref{mlcol.eq} was determined from the observed
relation between rest-frame $g-r$ color
and $M/L_g$ ratio for galaxies 
with $0.045<z<0.055$, $10<\log(M/M_{\odot}) < 10.7$, and
$0.2< (g-r) <1.2$
in the Sloan Digital Sky Survey DR7 (as provided by
the MPA-JHU release; {Brinchmann} {et~al.} 2004). 
The MPA-JHU relation between $M/L_g$ and $g-r$ has a scatter of 0.12\,dex
and assumes a {Chabrier} (2003) IMF. Instead of the observed colors,
which have large uncertainties at $R>20$\,kpc, we used a fit
of the form
\begin{equation}
(g-r) = \begin{cases} -0.32\log(R)+0.67, & \mbox{if }R\leq 29\,\mbox{kpc} \\
   0.20, & \mbox{if }R>29\,\mbox{kpc.} \end{cases}
\end{equation}
This fit is indicated by the broken green line in Fig.\ 2b.

The mass density profile is shown in Fig.\ \ref{massdens.fig}.
The form is similar to the surface brightness profile, except for
the central regions as those are more prominent in mass than
in blue light. The data reach surface densities of $\lesssim
10^4$\,\msun\,kpc$^{-2}$, and a non-parametric
limit on the stellar halo of M101 is that it has a stellar
mass density lower than this limit at $R\gtrsim 50$\,kpc.

\subsection{Fitting}

The profile is well-described by an exponential disk and 
a {Sersic} (1968) bulge. This
fit has the form
\begin{equation}
\rho(R) = \rho_{0,d} \exp\left(\frac{R}{R_d}\right) + \rho_{0,b}
\exp\left[-4.85 \left(\frac{R}{R_e}\right)^{1/n}\right],
\end{equation}
with $\rho_{0,d} = (4.40\pm 0.11) \times 10^8$\,\msun\,kpc$^{-2}$,
$R_d = 3.98 \pm 0.06$\,kpc, $\rho_{0,b} = (2.24 \pm 0.08) \times
10^{10}$\,\msun\,kpc$^{-2}$, $R_e = 1.67 \pm 0.12$\,kpc, and
$n=2.62 \pm 0.16$. The fit
is shown by the red solid line in Fig.\ \ref{massdens.fig}a.
We note that the ``bulge'' may in fact be more appropriately called
an inner disk; as is well known M101 has a very low central
velocity dispersion and spiral arms that continue
into the central few arcsec ({Kormendy} {et~al.} 2010). 

The residuals from the fit are shown below panel a of Fig.\
\ref{massdens.fig}. They are $<0.1$ dex at $R=0-40$\,kpc and
within the $1\sigma$ error bars at larger radii, confirming that
there is no significant upturn in the profile. We quantify the
contribution of a halo component by fitting the residuals.
To parameterize the halo we adopt model ``U'' in
the {Courteau} {et~al.} (2011) analysis of the M31 light profile
(their preferred model). This model is a power law:
\begin{equation}
\label{halo.eq}
\rho(R) = \rho_{0,h} \left[ \frac{1+(30/a_h)^2}{1+(R/a_h)^2}\right]^{\alpha}.
\end{equation}
The values of $a_h$ and $\alpha$ are fixed to the best fits for M31,
$a_h = 5.20$\,kpc and $\alpha=1.26$ (see Table 4 of {Courteau} {et~al.} 2011).
Fitting the normalization (i.e., the halo surface density at
30\,kpc) gives $\rho_{0,h} = 7^{+13}_{-7} \times 10^{3}$\,\msun\,kpc$^{-2}$.
The combined disk + bulge + halo model is shown by the red line in
Fig.\ \ref{massdens.fig}b.

The total mass implied by this model of M101, integrated to $R=200$\,kpc,
is $M_{\rm tot,*}=
5.3^{+1.7}_{-1.3}\times 10^{10}$\,\msun.
The halo mass is $M_{\rm halo}=
1.7^{+3.4}_{-1.7}\times 10^{8}$\,\msun, and we infer that the fraction
of mass in the halo is $f_{\rm halo} = M_{\rm halo} / M_{\rm tot,*}
= 0.003^{+0.006}_{-0.003}$. This fraction is significantly lower than
the halo fraction of M31: {Courteau} {et~al.} (2011) find $f_{\rm halo}
\sim 0.04$ using
the same decomposition method. We illustrate this difference in
Fig.\ \ref{massdens.fig}c, where we show what M101's profile
would look like if the galaxy had a 4\,\%, M31-like halo.
Such halos are clearly inconsistent with the data.

\section{Discussion}

We have measured the surface brightness profile of M101 to
$\sim 18$ disk scale lengths and to surface brightness levels
$\mu_g \sim 32$\,mag\,arcsec$^{-2}$. We 
do not find evidence for the presence of a stellar halo,
or more precisely for a photometric component at large
radii that can be distinguished from the disk.
Taking the halo profile of M31 as a model and fitting the normalization,
we find a halo fraction of $f_{\rm halo}=0.003^{+0.006}_{-0.003}$.

\begin{figure}[htbp]
\epsfxsize=8.6cm
\epsffile[28 426 340 717]{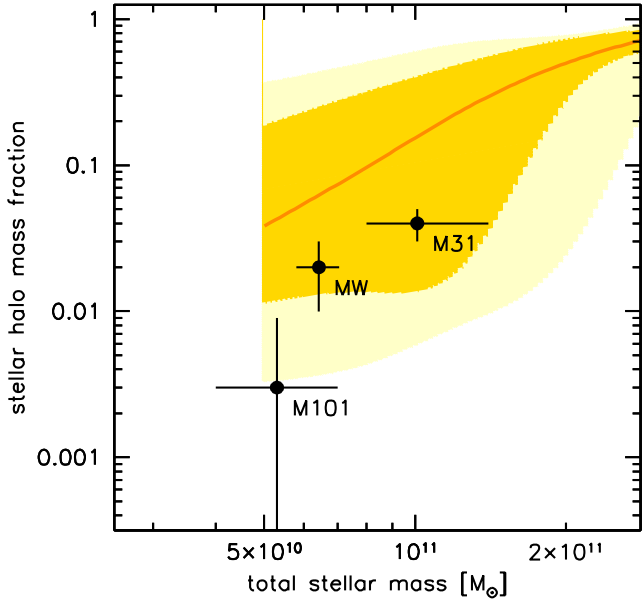}
\caption{\small
The mass fraction in the stellar halo as a function of the total
stellar mass. The stellar halo of M101
has significantly lower mass than those of the Milky Way and M31.
The orange line is the predicted median
relation between the accreted mass fraction and the total
stellar mass from numerical
simulations (Cooper et al.\ 2013; see text).
The yellow and brightyellow regions
indicate the 68\,\% and 95\,\% galaxy-to-galaxy variation
in the simulations.
\label{halofrac.fig}}
\end{figure}

This fraction is lower than that of M31 ($f_{\rm halo} \sim 4$\,\%;
Courteau et al.\ 2011) and also the Milky Way ($f_{\rm halo} \sim 2$\,\%;
Carollo et al.\ 2010; Courteau et al.\ 2011).
In Fig.\ \ref{halofrac.fig} we show the relation between
stellar halo fraction and galaxy stellar mass.
The stellar masses of the Milky Way 
and M31 were taken from {McMillan} (2011) and {Tamm} {et~al.} (2012)
respectively. For comparison, we
show the relation between the accreted
fraction of stars and galaxy stellar
mass as
predicted by numerical models that trace the light of accreted
satellites in dark matter halos. This relation
was derived from the data in Fig.\ 12b of Cooper et al.\ (2013)
for bulge-to-total ratios $B/T<0.9$; the relation for other
$B/T$ limits is very similar.


Interestingly, the stellar halo masses of all three galaxies are below
the predicted relation, with M101 a factor of $\sim 10$
below the median expectation.
We caution, however, that stellar
halo masses are not measured in a self-consistent way in such comparisons.
There is no universal definition of a stellar halo,
and it is unclear whether it even makes sense to model
it as a single component
(see, e.g., {Carollo} {et~al.} 2010). From a practical perspective it
is perhaps most fruitful to test the model predictions
by comparing the predicted and observed
radial surface density profiles directly,
or by fitting
a model such as Eq.\ \ref{halo.eq} with
only the normalization as a free parameter. In this context it is
interesting to note that the {Cooper} {et~al.} (2010, 2013) models
predict that stellar halos begin to dominate at $R\sim 20$\,kpc
and surface densities $\rho \sim 10^5$\,\msun\,kpc$^{-2}$ -- again
inconsistent with the M101 observations (but not with M31).

Given the stochastic nature of accretion events and, as a consequence,
the large scatter predicted in the $f_{\rm halo} - M_{\rm tot,*}$
relation ({Purcell} {et~al.} 2008; {Cooper} {et~al.} 2013; Fig.\
4), it is important to build up
a sample of galaxies with radial profiles reaching surface densities
of $\sim 10^4$\,\msun\,kpc$^{-2}$.
Star counts with the Gemini telescope ({Bland-Hawthorn} {et~al.} 2005), the Subaru
telescope ({Tanaka} {et~al.} 2011) and the Hubble Space
Telescope ({Barker} {et~al.} 2009; {Radburn-Smith} {et~al.} 2011; {Monachesi} {et~al.} 2013) have reached depths of
$\gtrsim 30$\,mag\,arcsec$^{-2}$  but
only for very nearby, low mass galaxies. Reaching those limits
at distances beyond $\sim 5$\,Mpc is exceedingly difficult
as the apparent brightness of stars decreases
with the square of their distance.
By contrast, the integrated-light surface
brightness is independent of distance, and low surface brightness-optimized
telescopes such as
Dragonfly can study galaxies out to the Virgo cluster and
beyond.\footnote{{Bland-Hawthorn} {et~al.} (2005) have
pointed out that scattered light off dust grains can contribute to
integrated-light measurements at very faint levels for very compact
galaxies; this should not be a concern for M101.}
This makes it possible to construct complete samples and
search for correlations of $f_{\rm halo}$ with other galaxy parameters.
As an example, elliptical galaxies and spiral galaxies
likely had different accretion histories
(e.g., {Guedes} {et~al.} 2011; {Tal} \& {van Dokkum} 2011; {Cooper} {et~al.}
2013; {van Dokkum} {et~al.} 2013).

\begin{figure}[htbp]
\epsfxsize=8.6cm
\epsffile[0 0 633 637]{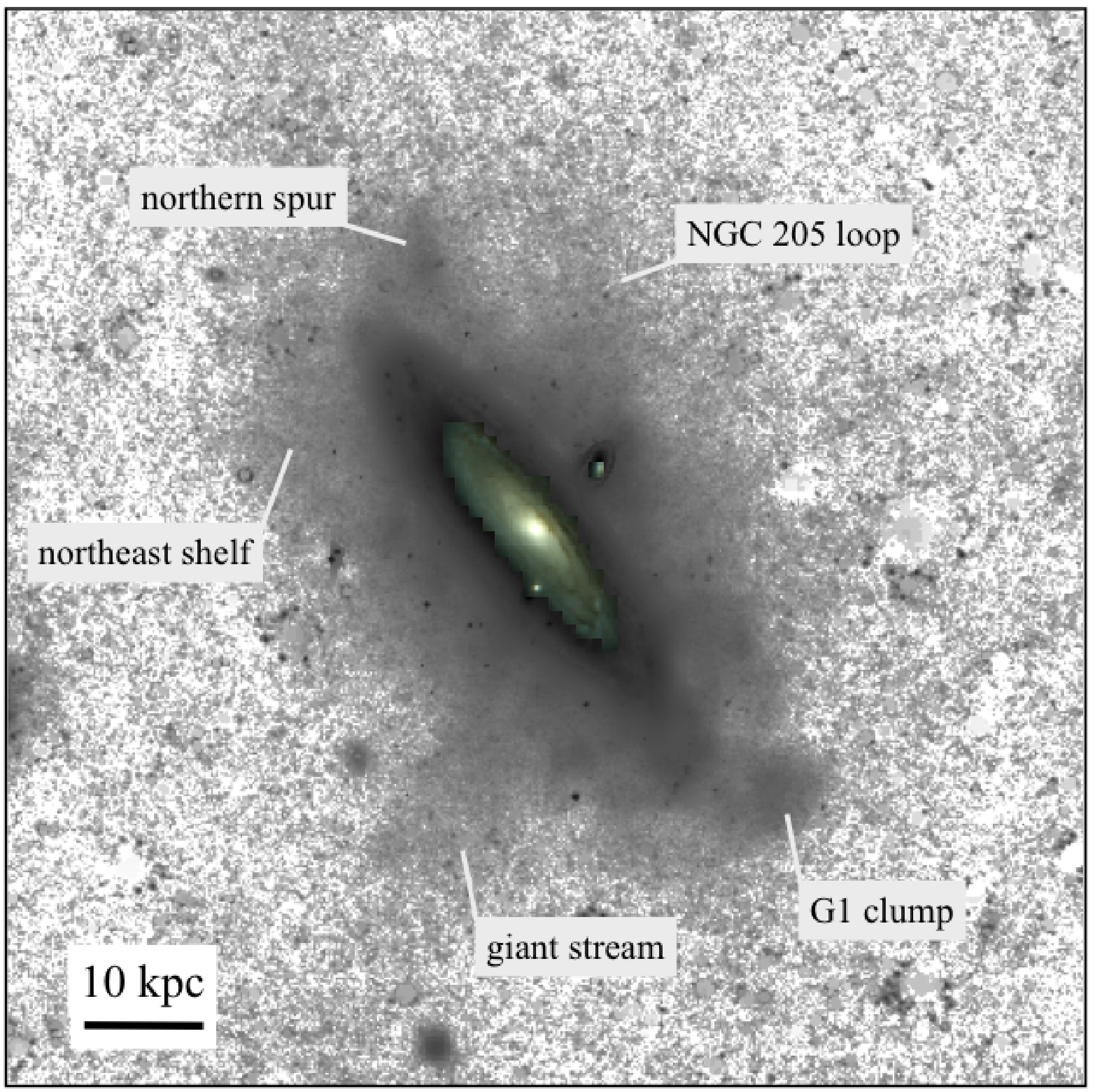}
\caption{\small
M31 redshifted to 7 Mpc and placed in an empty region of our
M101 image. The image has the exact same scaling as Fig.\ 1c.
Well-known features in the M31 halo are labeled; they would be
easily detected with Dragonfly.
\vspace{0.5cm}
\label{m31.fig}}
\end{figure}

Finally, we note that halos can also be identified by their substructure,
as it reflects the detailed accretion history of a galaxy
(see, e.g., {Fardal} {et~al.} 2008, for a discussion on M31). This
history is not very informative for an individual galaxy but a
large sample can
provide very strong constraints on galaxy formation models
({Johnston} {et~al.} 2008). To illustrate
the capabilities of the Dragonfly Telephoto Array in this context,
we created a simulated Dragonfly image of M31 by redshifting
this galaxy to the distance of M101 (Fig.\ \ref{m31.fig}).

The M31 observations that were used are a combination of
a Dragonfly image taken
on 26 June 2013 and star count data from PAndAS
({McConnachie} {et~al.} 2009; {Carlberg} {et~al.} 2011). The star counts go out very far from
the center of M31 
but they are incomplete at small radii due to crowding.
Following a similar procedure to that
described in {Irwin} {et~al.} (2005)
Dragonfly data at
$R>0.7^{\circ}$ were used to tie the star count data to the integrated-light
data.
The combined Dragonfly + PAndAS image was
redshifted to 7.0\,Mpc and
placed in a relatively empty region of
the full-field M101 image (see Fig.\ \ref{full.fig}b). This
last step ensures that the noise characteristics and artifacts from the
reduction are identical to the M101 data. As the scaling is
identical to Fig.\ \ref{full.fig}c the actual Dragonfly image
of M101 can be compared directly to the simulated Dragonfly
image of M31.

Prominent features in the M31 halo (see Fig.\ 1
in {Ferguson} {et~al.} 2005) are labeled in Fig.\ \ref{m31.fig}.
Remarkably, these features are all clearly visible,
including the famous giant stream first identified by {Ibata} {et~al.} (2001).
It has been known for many decades that dramatic tidal features
can be
detected in integrated light; telescopes such as Dragonfly are
now enabling us
to detect the subtle relics of galaxy formation that
should be present around every $\gtrsim L_*$ galaxy.

\begin{acknowledgements}
We are grateful to the PAndAS team for sharing their
M31 star count data in digital form, to Andrew Cooper for
providing us with the data to construct the model curves
in Fig.\ 4, and to the staff at New
Mexico Skies for their dedication and support. The anonymous
referee is thanked for excellent comments that substantially
improved the manuscript.  We thank
the NSF (grant AST-1312376) and NSERC for financial support.
\end{acknowledgements}


\end{document}